\documentclass[runinaddress,showpacs,aps,prb,superscriptaddress,a4paper,preprint,floatfix]{revtex4-1}
\usepackage{amsfonts}
\usepackage{amsmath}
\usepackage{amssymb}
\usepackage{bm}
\usepackage{color}
\usepackage{color,graphicx}
\usepackage{dcolumn}	% Align Table  columns on decimal point
\usepackage{gensymb}
\usepackage{graphicx}	% Include Figure files
\usepackage[hyperfootnotes,hyperindex,hyperfigures,pagebackref=false]{hyperref}%{pagebackref}
\usepackage{latexsym}
\usepackage{mathcomp}
\usepackage{textcomp}
\usepackage{verbatim}
\usepackage{dcolumn}
\usepackage{ulem}
\providecommand{\by}{$\times$}

\providecommand{\STO}{SrTiO$_{3}$}

\providecommand{\LAO}{LaAlO$_{3}$}

\providecommand{\alo}{AlO$_6$~}

\providecommand{\Vn}[1]{$V_{\mathrm{#1}}$}

\definecolor{red}{rgb}{0.8,0,0.2}

\def\beeq{\begin{equation}}
\def\eneq{\end{equation}}
\def\beeqa{\begin{eqnarray}}
\def\eneqa{\end{eqnarray}}

\def\FE{\textcolor{red}}
\definecolor{adobe}{rgb}{.8,.6,.5}
\definecolor{mygreen}{rgb}{0.0,0.6,0.0}
\definecolor{blue2}{rgb}{0.0,0.0,0.8}
\definecolor{brown}{rgb}{0.6,0.3,0.0}
\definecolor{forest}{rgb}{0.0,0.4,0.0}
\definecolor{grass}{rgb}{0.0,0.55,0.25}
\definecolor{grass2}{rgb}{0.0,0.6,0.25}
\definecolor{gray}{rgb}{0.4,0.4,0.4}
\definecolor{grayish}{rgb}{0.2,0.2,0.4}
\definecolor{khaki}{rgb}{0.9,0.9,0.7}
\definecolor{lightteal}{rgb}{0.0,0.6,0.6}
\definecolor{lightyellow}{rgb}{1.0,1.0,0.5}
\definecolor{maroon}{rgb}{0.7,0.1,0.2}
\definecolor{navy}{rgb}{0.0,0.1,0.7}
\definecolor{olive}{rgb}{0.4,0.4,0.0}
\definecolor{orange}{rgb}{0.9,0.45,0.0}
\definecolor{peach}{rgb}{1.0,.8,.7}
\definecolor{purple}{rgb}{0.4,0,0.55}
\definecolor{teal}{rgb}{0.0,0.5,0.4}
\definecolor{turq}{rgb}{0.3,0.6,0.9}
\definecolor{violet}{rgb}{0.75,0,0.75}
\hypersetup{colorlinks=true, linkcolor=navy, citecolor=blue, urlcolor=blue, filecolor=red, anchorcolor=blue}% linkcolor=FIGURE numbers citecolor=citation colors

\def\FE{\textcolor{black}}
\begin{document}
\DeclareGraphicsExtensions{.ps,.pdf,.eps,png}

\preprint{UPDATED: {\color{maroon} \today}}

%: TITLE OF PAPER +++++++++++++++++++++++++++++++++++++++++++++++++++++++++++
\title{{\color{navy}The role of screened exact exchange in accurately describing properties of transition metal oxides:  Modeling defects in \LAO}}

%: AUTHOR LIST ++++++++++++++++++++++++++++++++++++++++++++++++++++++++++++++++
\author{\firstname{Fedwa} \surname{El-Mellouhi}}
   \email{fadwa.el\_mellouhi@qatar.tamu.edu}
     \affiliation{Chemistry Department, Texas A\&M at Qatar, Texas A\&M Engineering Building, Education City, Doha, Qatar}
     \affiliation{Physics Department, Texas A\&M at Qatar, Texas A\&M Engineering Building, Education City, Doha, Qatar}

\author{\firstname{Edward N.} \surname{Brothers}}
   \email{ed.brothers@qatar.tamu.edu}
     \affiliation{Chemistry Department, Texas A\&M at Qatar, Texas A\&M Engineering Building, Education City, Doha, Qatar}

\author{\firstname{Melissa J.} \surname{Lucero}}
  \affiliation{Department of Chemistry, Rice University, Houston, Texas 77005-1892}

\author{Gustavo E. Scuseria}
  \affiliation{Department of Chemistry, Rice University, Houston, Texas 77005-1892}
  \affiliation{Department of Physics and Astronomy, Rice University, Houston, Texas 77005-1892}
 \affiliation{Chemistry Department, Faculty of Science, King Abdulaziz University, Jeddah 21589, Saudi Arabia }
%\email{Electronic mail: guscus@rice.edu}
%\homepage{http://python.rice.edu/~guscus/}

%: ABSTRACT ++++++++++++++++++++++++++++++++++++++++++++++++++++++++++++++++  
\begin{abstract}

The properties of many intrinsic defects in the wide band gap semiconductor
\LAO~are studied using the screened hybrid functional of Heyd, Scuseria, and
Ernzerhof (HSE). As in pristine structures, exact exchange included in the screened
hybrid functional alleviates the band gap underestimation problem, which is common to semilocal
functionals; this allows accurate prediction of defect properties.  We propose
correction-free defect energy levels for bulk  \LAO~computed using HSE that
might serve as guide in the interpretation of photoluminescence experiments.

\end{abstract}

\pacs{71.15.Mb,% Density functional theory, local density approximation , gradient and other corrections.
 71.15.Ap,% Basis sets (LCAO, plane-wave, APW, etc.) and related methodology (scattering methods, ASA, linearized methods, etc.)} %  Dielectric, piezoelectric, ferroelectric, and antiferroelectric materials
}

\maketitle

Defects in \LAO~have been studied extensively both
experimentally~\cite{Chen:2011} and using computational
approaches,~\cite{Luo:2009,Yamamoto:2012} contributing to our understanding of
the interplay between various defects in this material. Photoluminescence (PL)
spectroscopy using sub-band-gap excitation  was recently used to detect the
\FE{ground state} defect states within the band gap of \LAO~single
crystals.\cite{Chen:2011}  \FE{In standard  photoluminescence,  electrons are pumped to the conduction
band then a photon is emitted upon relaxation  from conduction band to various ground state defect levels. The resulting   PL peaks are then  associated with defect levels inside the gap. In sub-bandgap excitation, the photon energy is tuned to selectively probe certain defect levels revealing  more detailed  features.} This experiment identified three distinct PL peaks,
each showing doublet splitting, that  were localized 2~eV below the conduction
band minimum (CBM). Defect levels calculated~\cite{Luo:2009} using the
the generalized gradient approximation density functional of Perdew, Burke and
Ernzerhof (PBE)\cite{Perdew:1996,Perdew:1997} and corrected with the
``scissor operator''  were used as a guideline to partially match the PL peaks.
This approach is less than completely satisfying, however, as (for example) the
La$_{\mathrm Al}$ defect level,  post-correction, is located  1~eV below  the
CBM; this contradicts recent experimental results.  A more accurate theoretical
description is thus much needed, especially given the problems of band gap
underestimation (endemic to semilocal functionals)~\cite{Heyd:2005bh}  which is
fatal for defect calculations,  and questions about the overall appropriateness
of the ``scissor operator.'' Put more simply, the typical theoretical methods
which can be used for modeling these sorts of materials are insufficiently
accurate for explaining the effects in question. 

Defects in \LAO~have been subject to other very recent theoretical
calculations:~\cite{Yamamoto:2012, Mitra:2012} Vacancy defect energetics in
rhombohedral  and cubic bulk \LAO~ have been  computed using PBE in
Ref.~\onlinecite{Yamamoto:2012}, where it was found that the defect formation behavior
in both phases were very similar. That work also included finite size scaling
using supercells up to 480 atoms, suggesting that the cell-size dependencies in
modeling neutral vacancies are almost negligible. (This makes their formation
energies almost independent from the supercell size.) However, it should be
noted that formation energies were modified using a band-gap correction
scheme~\cite{Yamamoto:2012} to overcome the well-known band gap underestimation
problem of semilocal functionals. For this reason, interest has emerged in using
modern (and demonstrably more accurate~\cite{Henderson:2011}) screened hybrid
functionals to model these defects. While some recent efforts have been published in
this direction,~\cite{Xiong:2008, Mitra:2012} a complete picture of all possible
defect levels using modern hybrid functionals is not available. 

In the present work, we apply the screened hybrid functional of
Heyd-Scuseria-Ernzerhof (HSE) to a wide array of neutral defect types in \LAO,
thus complementing  previous HSE efforts~\cite{Mitra:2012} that  treated only
the oxygen vacancies.  This work is motivated by HSE's agreement with experiment
for the calculation of many of the electronic, structural, and elastic properties
in cubic \LAO.~\cite{El-Mellouhi:2013} HSE is expected to  give point defect
formation energies and energy levels in close agreement with experiment as its
direct and indirect band gaps~\cite{El-Mellouhi:2013} as well as valence band
widths (VBW)\cite{Ramprasad:2012} are in excellent agreement with experiment (see
table~\ref{tab:enthalpies}); this can be contrasted with the PBE results, which have
been  previously used to study point defects in \LAO.~\cite{Luo:2009,
Yamamoto:2012} 
\FE{It is worth noting that HSE06 gives an excellent agreement with the results of the global hybrid PBE0 for the case of the oxygen vacancy in \STO~\cite{Evarestov:2012}. This suggests that hybrid functionals belonging to the 25\% HF exchange family  like PBE0 and HSE06 would yield very similar location of the defect level and the splitting of the conduction band minimum in the \LAO~case as well.}

Here we restrict our study to neutral defects  to  avoid introducing errors due
to spurious electrostatic interactions,  and the corrections associated with it.
Nevertheless, performing HSE calculations with the high numerical accuracy
settings detailed below remains quite expensive,  thus precluding the use of the
largest supercells.  This is acceptable, however, as finite size scaling and
previous investigations~\cite{ Mitra:2012, Yamamoto:2012} using larger
supercells have shown that the neutral defects considered here suffer least from
finite size effects. Consequently, despite the limited number of atoms that can
be treated with HSE, this approach promises  increased physical accuracy
compared to the less expensive semilocal functionals.

All calculations presented in this paper were performed using the development
version of the {\sc gaussian} suite of programs,\cite{gdv} with the periodic
boundary condition (PBC)\cite{Kudin:2000} code used throughout.  The
Def2-\cite{Weigend:2005} series of Gaussian basis sets were optimized following
our procedure described in Ref.~\onlinecite{El-Mellouhi:2011} for bulk \LAO.
As in Ref.~\onlinecite{El-Mellouhi:2011}, we use the notation SZVP  to
differentiate these optimized PBC basis sets from the molecular Def2-SZVP basis
sets.  The functionals applied in this work include
PBE\cite{Perdew:1996,Perdew:1997} and HSE.\cite{HSEh} 

% ED says:  I don't think we need this, as it is covered by the reference quite well.
%Briefly, the HSE functional partitions the coulomb potential into short-range (SR) and long-range (LR) components:
%\begin{equation}
%E^{HSE}_{xc}= \frac{1}{4}E^{HF, SR}_{x}(\omega) + \frac{3}{4}E^{PBE, SR}_{x}(\omega) 
%+ E^{PBE, LR}_{x}(\omega) + E^{PBE}_{c}
%\end{equation}

%The screening parameter $\omega_{HF}$=$\omega_{PBE}$=0.11 $bohr^{-1}$ defines the  separation range, as it controls
%the distance at which the long-range nonlocal interaction becomes negligible, {\it i.e.} it ``turns off'' exact exchange after a specified distance.  

 The use of HSE imposes limitations on the
size of the supercell that can be efficiently computed \FE{and fully relaxed}, so a LAO supercell of
2$\times$2$\times$2 \FE{replica of the 5 atoms cubic unit cell } (40 atoms) was used with a dense $k$-point mesh of
6$\times$6$\times$6,  including the $\Gamma$ point. Also, we modeled a larger
supercell of  2$\times$3$\times$3 (90 atoms), with the same density of
$k$-points, in order to discuss the importance of defect self-interactions, and
the effect of varying the defect concentration on the electronic properties of
LAO. 

Most numerical settings in {\sc gaussian} were left at the default values,
\textit {e.g.}, geometry optimization settings, integral cut-offs, $k$-point
meshes and SCF convergence thresholds.  Unless otherwise noted, crystal
structures used in the chemical potential calculations on La, Al, Al$_\mathrm
2$O$_\mathrm 3$, La$_\mathrm 2$O$_\mathrm 3$ were downloaded as CIF files from
the ICSD,\cite{ICSDLAO}  and then fully relaxed/optimized.  \FE{Isolated, neutral
vacancies were introduced to the crystal structure of cubic LAO by removing one
atom of either O, La or Al, while La and Al antisites occupied the crystalline position. All
structures containing the above defects were then fully relaxed using HSE06.  In order to avoid imposing a certain oxygen interstitial position, the oxygen atom was inserted  far from the well-known interstitial sites followed with relaxation to the nearest minimum. At this point, we cannot be completely sure whether the configuration we obtained has the lowest formation energy; only a full energy landscape exploration method can reveal that. }

%%%%%%%%%%%%%%%%%%%%%%%%%%%%%%%%%%%%%%%%%%%%%%%%%%%%%%%%%%%%%%%%%%
\begin{table}[!htb]
\caption{\label{tab:cool} Comparison of calculated fundamental electronic
properties of bulk cubic \LAO~from this work and previous studies. VBW stands
for the valence band width. Calculated enthalpies of formation in eV/atom for
idealized materials with phases containing La, Al and O  are compared to previous PBE
calculations~\cite{Luo:2009} and experiment.  }
\begin{ruledtabular}
\begin{tabular}{lllllll}
\label{tab:enthalpies}
		&\multicolumn{2}{c}{This Work}  &\multicolumn{2}{c}{Previous Work}   \\
		&HSE &PBE &PBE &Exp.\\
\hline
\\
Direct gap (eV)   		&5.0 &3.54  &- &-\\
Indirect gap (eV)   	&4.74 &3.26 &3.1\footnotemark[5] &-\\
VBW (R$\rightarrow$R)(eV) &8.00 &7.50  &- &-	 \\
\\
\hline
\\
$\Delta$H$^\mathrm f_\mathrm{Al_2O_3}$	 	&3.82	&3.6 	&3.30\footnotemark[1]	 &3.47\footnotemark[2] \\	
\\					
$\Delta$H$^\mathrm f_\mathrm{La_2O_3}$ 		&4.24	&4.00	&3.71\footnotemark[1]	&3.71\footnotemark[3]  \\	\\					
$\Delta$H$^\mathrm f_\mathrm{LaAlO_3}$ 		&3.78	&4.21 	&3.60\footnotemark[1] 	&3.45\footnotemark[4] \\		\\				

\end{tabular}
\end{ruledtabular}
\footnotetext[1]{Ref. \onlinecite{Luo:2009}} 
\footnotetext[2]{Ref. \onlinecite{NIST:2008}}
\footnotetext[3]{Ref. \onlinecite{Cordfunke:2001}}
\footnotetext[4]{Ref. \onlinecite{Schnelle:2001}}
\footnotetext[5]{Ref~\onlinecite{Xiong:2008}}
\end{table}
%%%%%%%%%%%%%%%%%%%%%%%%%%%%%%%%%
\begin{figure}[!h]
\includegraphics[width=\columnwidth]{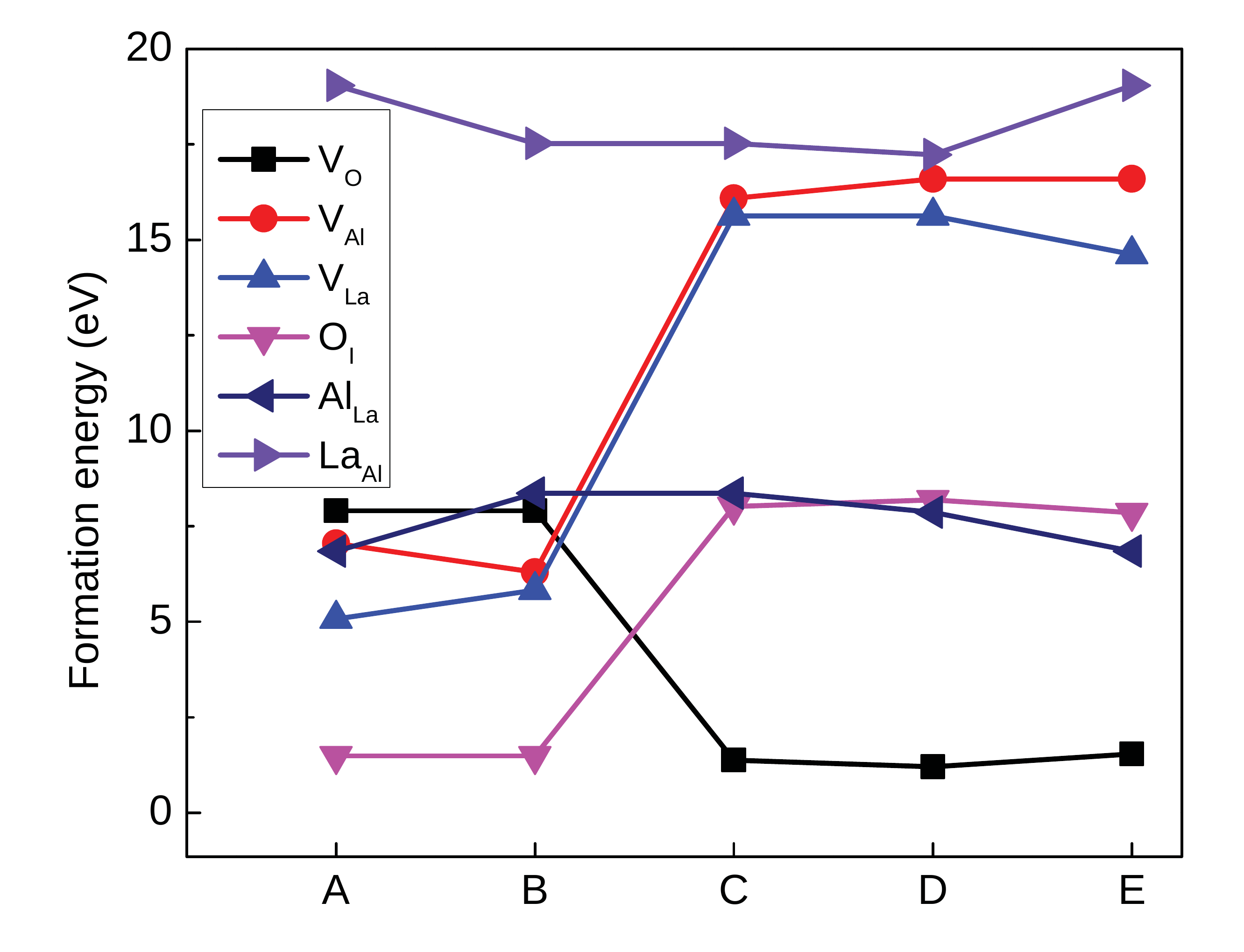}
\caption{\label{fig:formation}   Defect formation energies of isolated defects in cubic LAO computed using HSE at each equilibrium point based upon the phase diagram  in Ref.~\onlinecite{Luo:2009}.}
\end{figure}

The calculations of neutral defect formation energies used the formalism of Zhang and Northrup,\cite{Zhang:1993} namely the equation:
\begin{equation}\label{eqn:tanaka}
    \begin{split}
    E_{f} & = E_{T}-[E_{T}(\mathrm{perfect})\\
        & - ~n_{\scriptscriptstyle{La}}\mu_{\scriptscriptstyle{La}}-n_{\scriptscriptstyle{Al}}\mu_{\scriptscriptstyle{Al}}-n_{\scriptscriptstyle{O}}\mu_{\scriptscriptstyle{O}}]\\
    \end{split}    
\end{equation}

\noindent where $E_{T}$ and $E_{T}(\mathrm{perfect})$ are the calculated total
energies of the supercells containing the point defect and the perfect bulk host
materials, respectively.  The number of each element removed from the perfect
supercell is represented by $n_{\scriptscriptstyle{x}}$,  while
$\mu_{\scriptscriptstyle{x}}$ corresponds to the atomic chemical potentials in
an \LAO~ crystal.  Assuming that \LAO~is always stable, the chemical
potentials of the these elements can vary in the following correlation:

\begin{equation}
\mu_{La}+\mu_{Al}+3\mu_O =\mu^{bulk}_{LaAlO_3}
\end{equation}

Obviously, atomic chemical potentials are determined by the sample composition
and cannot be ascertained exactly.  However, they can be varied to cover the
whole phase diagram of LAO splitting into  Al$_2$O$_3$ and  La$_2$O$_3$ bulk
phases.  Hence the calculated  formation energies for the  neutral point defects
vary according to equilibrium positions such as ``O-rich'' and ``O-poor'' conditions.  

The calculated enthalpies of formation in idealized materials (non-relaxed
structures) for phases containing La, Al and O are summarized in
Table~\ref{tab:enthalpies}  and are compared to previous
calculations\cite{Luo:2009} and experiments. 
As a general trend, the formation enthalpies computed with HSE are close to  the
results from semilocal functionals  like PBE (this work),
although the HSE values are slightly higher. The only exception is LaAlO$_3$,
where  PBE tends to overestimate the formation enthalpies and  exceed the HSE
value. 
 
\begin{figure*}
\label{fig:bands}
\includegraphics[height=0.25\textheight]{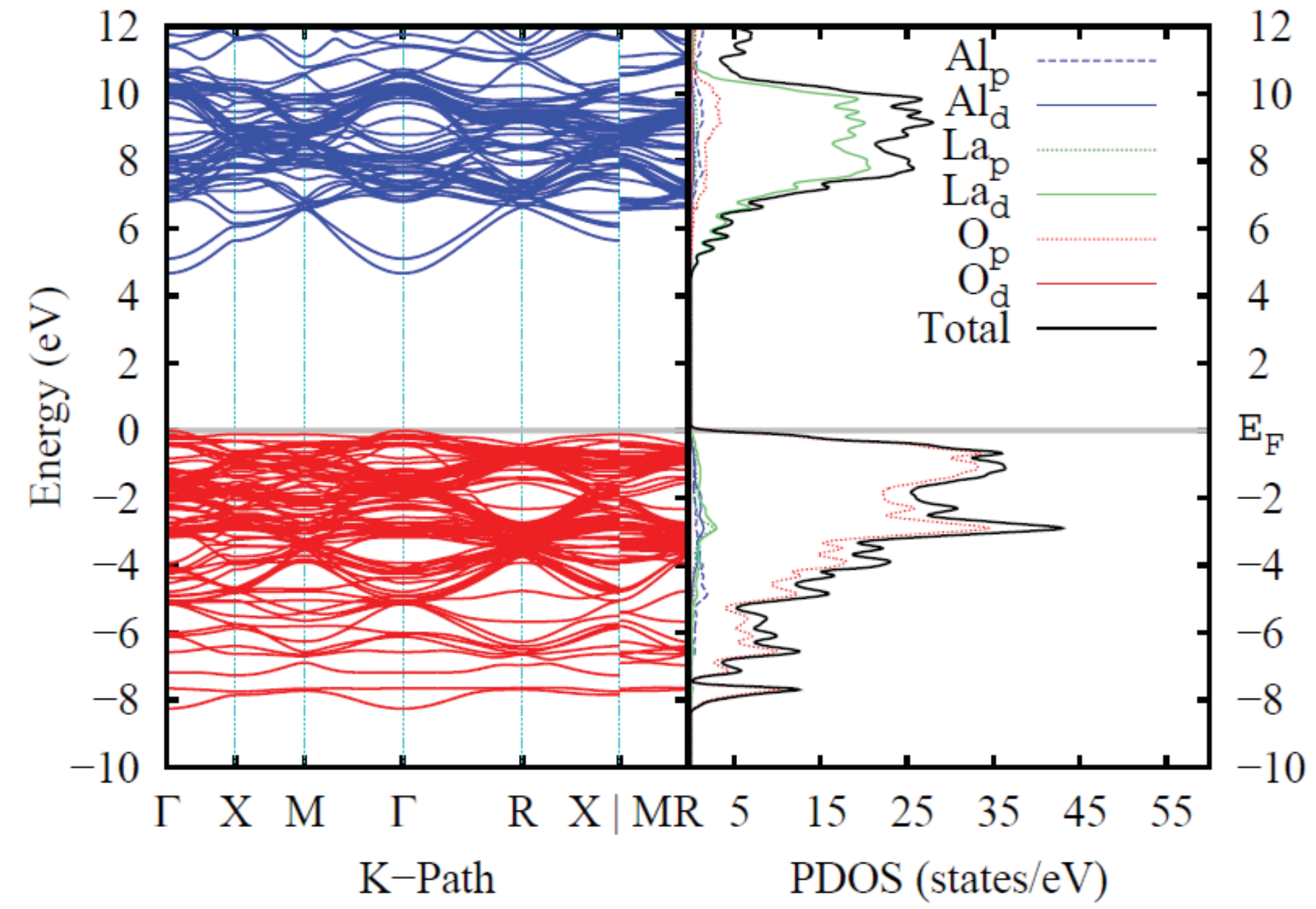}\includegraphics[height=0.25\textheight]{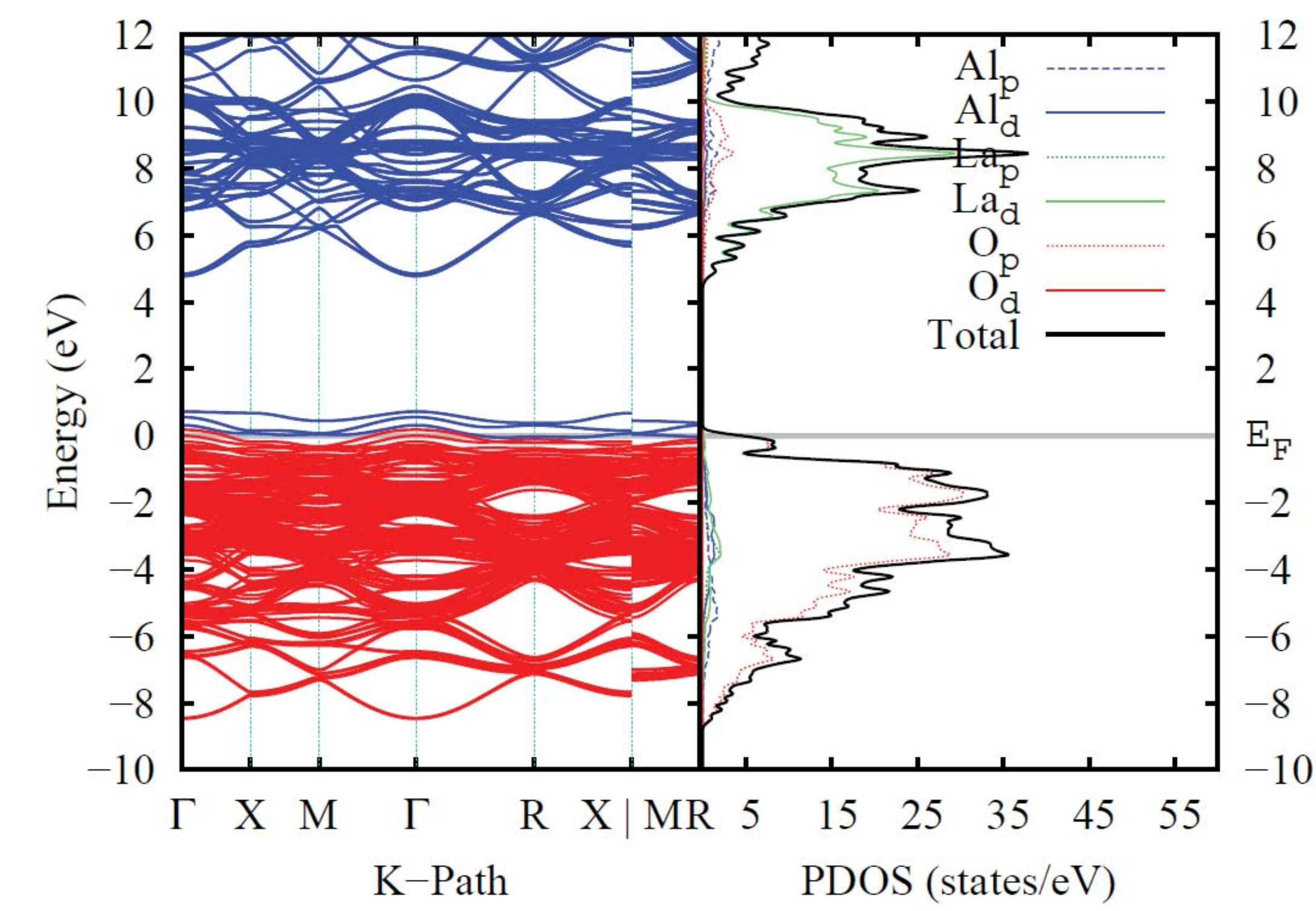}
\includegraphics[height=0.25\textheight]{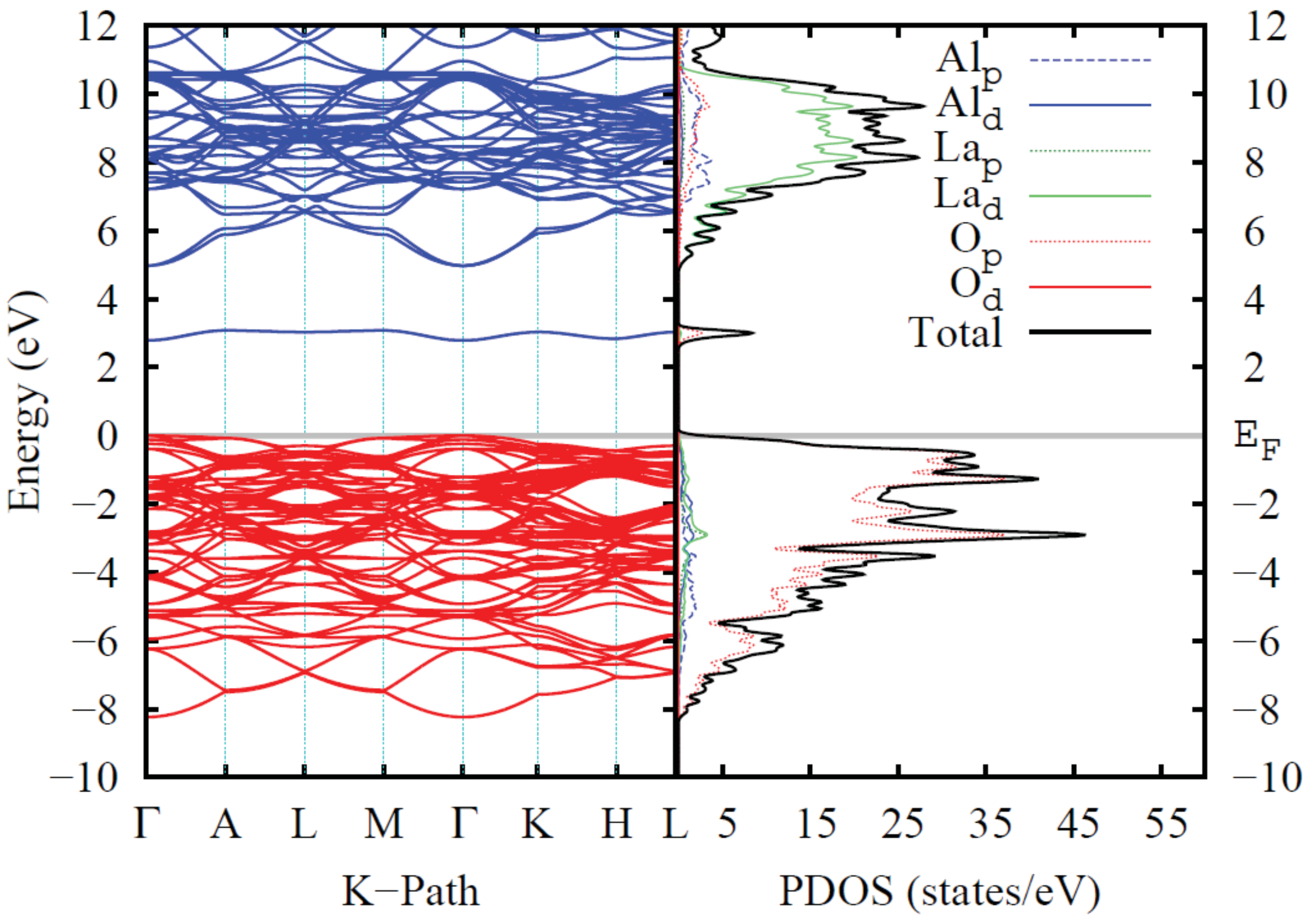}\includegraphics[height=0.25\textheight]{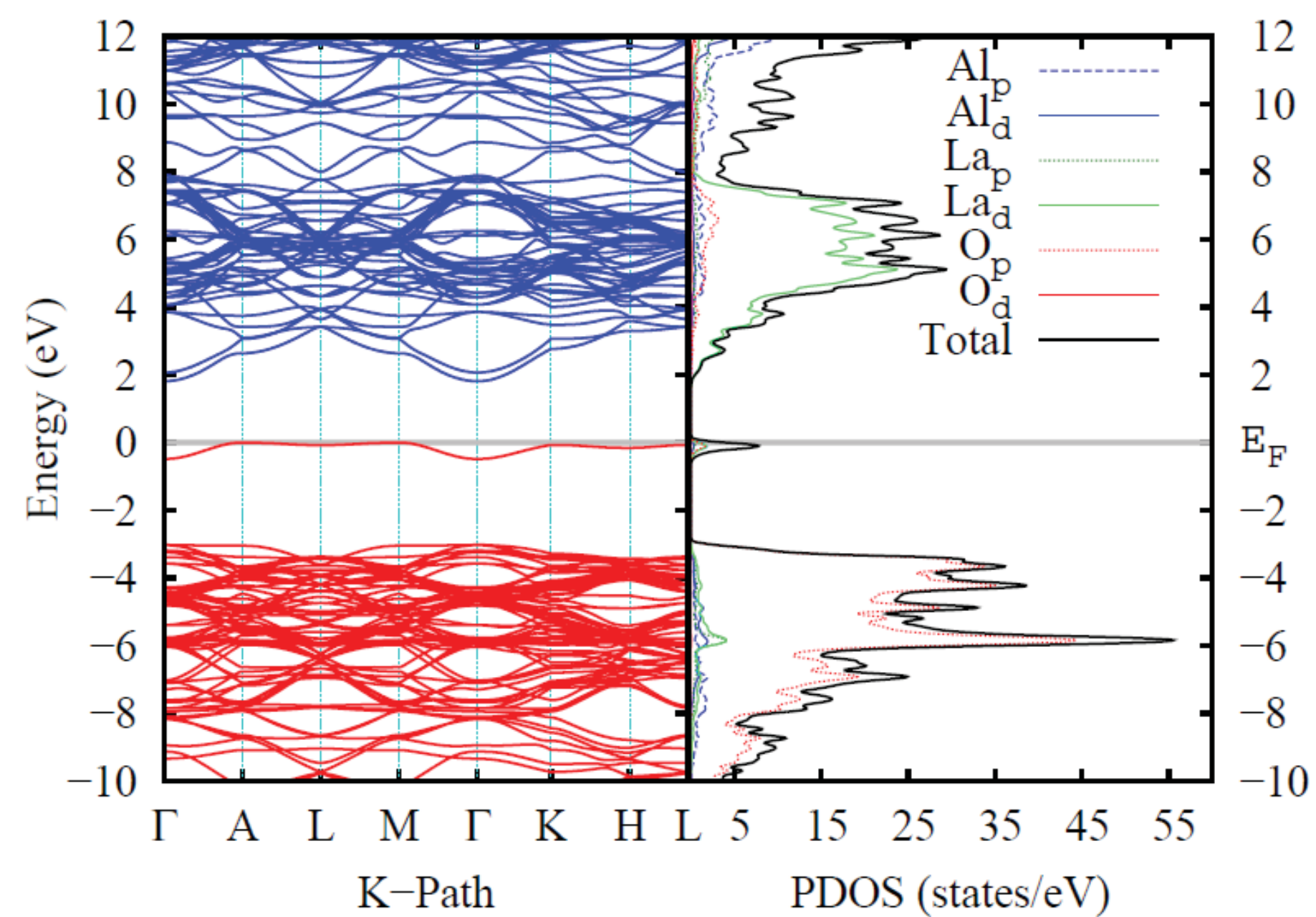}
\includegraphics[height=0.25\textheight]{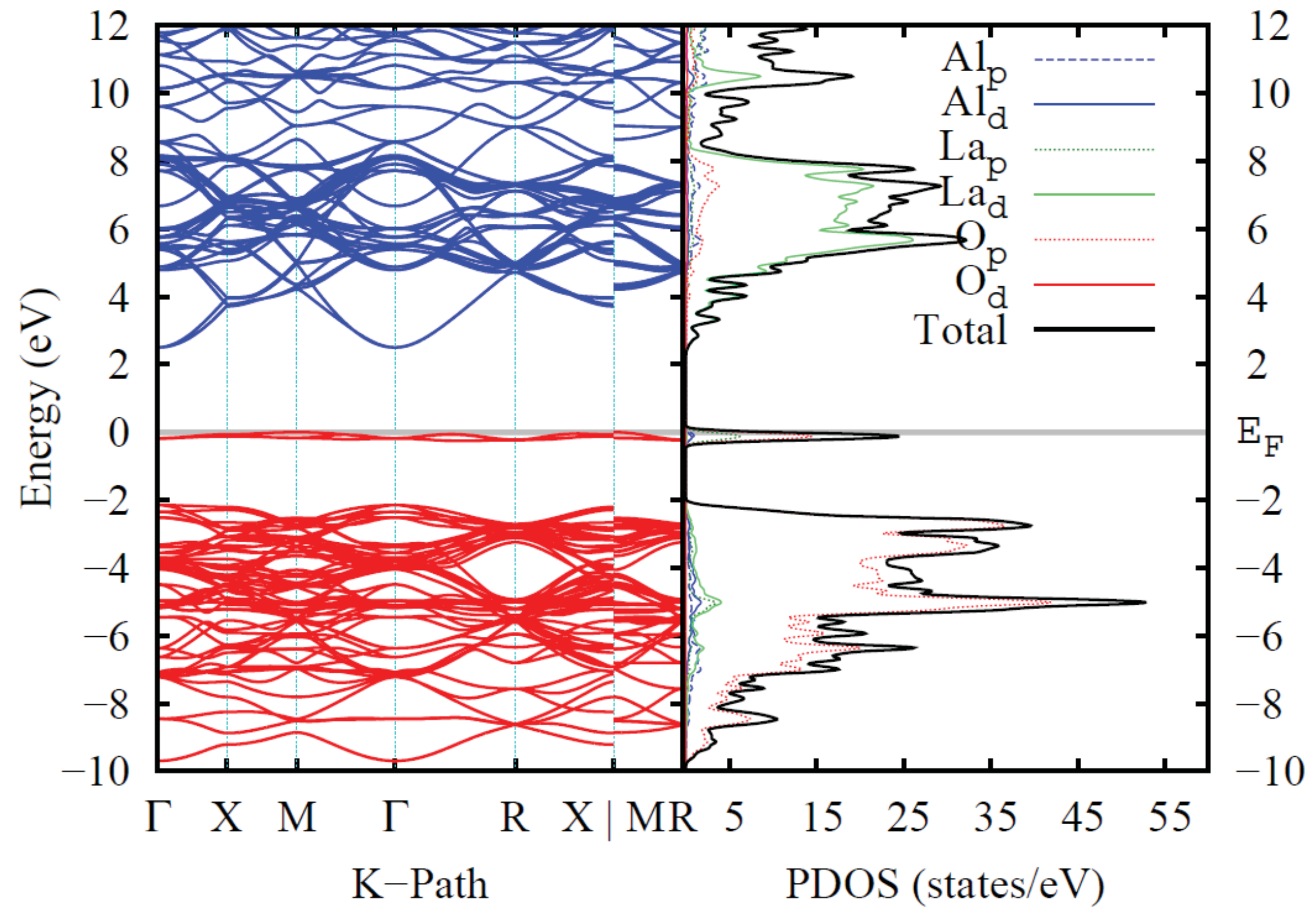}\includegraphics[height=0.25\textheight]{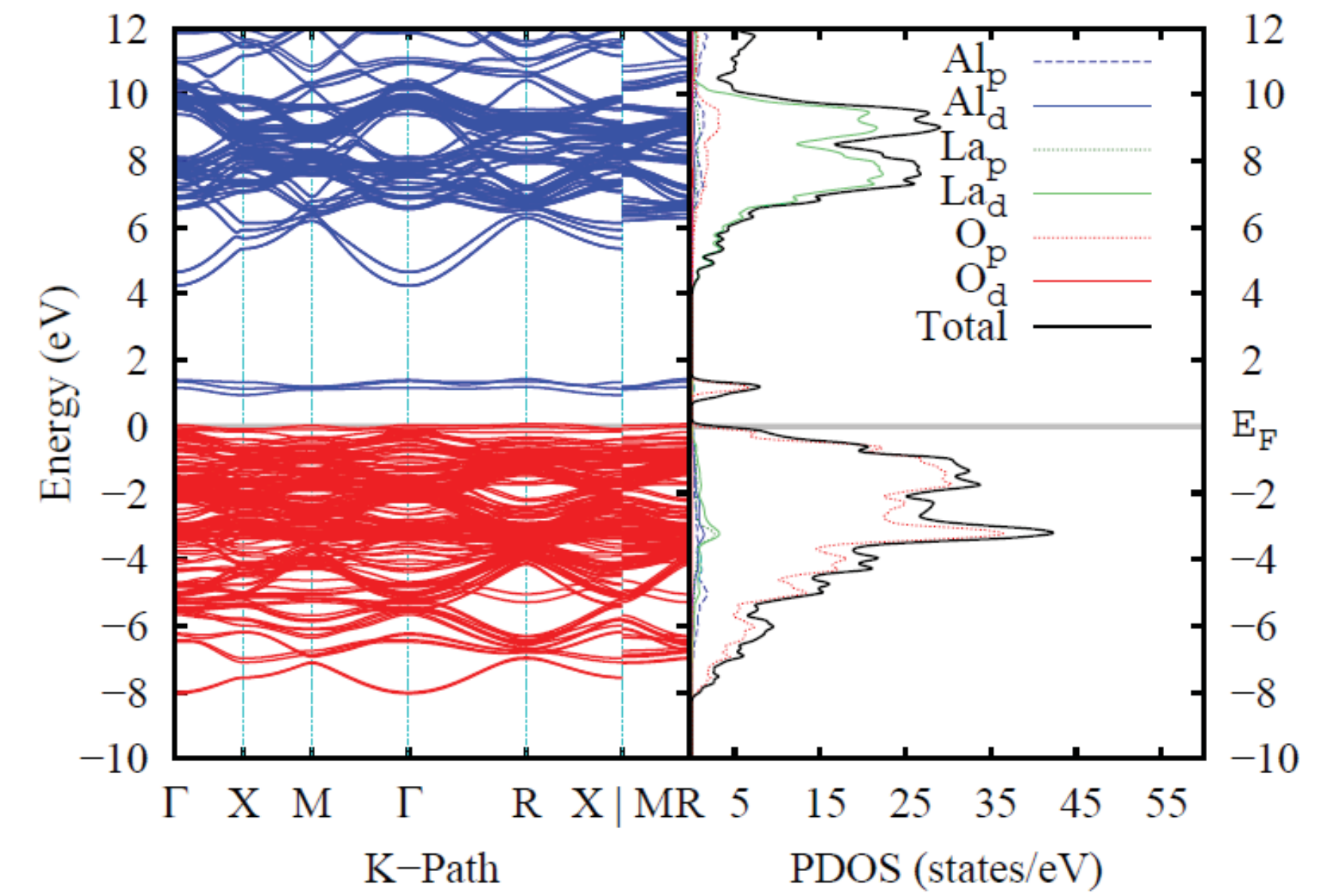}
\caption{\label{fig:bands} Band structures and PDOS calculated with HSE/SZVP for the 2\by2\by2 \LAO~supercell containing intrinsic defects. The top figures represent O$_{\mathrm I}$ and \Vn{La} introducing bands with a valence band character. Al$_{\mathrm La}$ and \Vn{O} (middle row) have bands  above the mid gap. The bottom row contains La$_{Al}$ and \Vn{Al} having defect bands below mid gap. The Fermi  energy \textit{E$_{\mathrm F}$} is indicated by a solid black line. The red bands indicate the occupied defect bands while the unoccupied defect bands are shown in blue. }
\end{figure*}

\begin{figure*}[]
\includegraphics[width=.5\columnwidth]{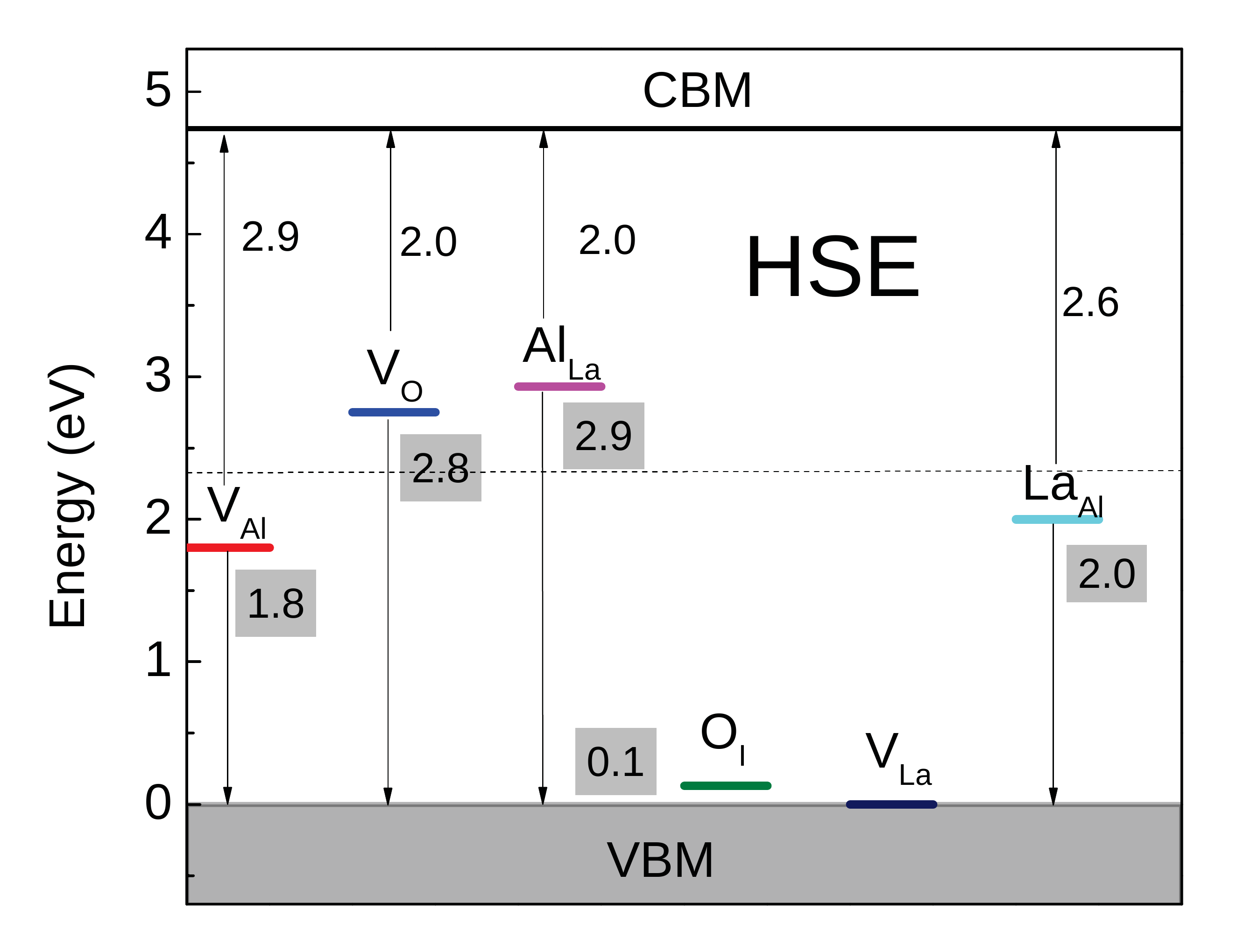}\includegraphics[width=.5\columnwidth]{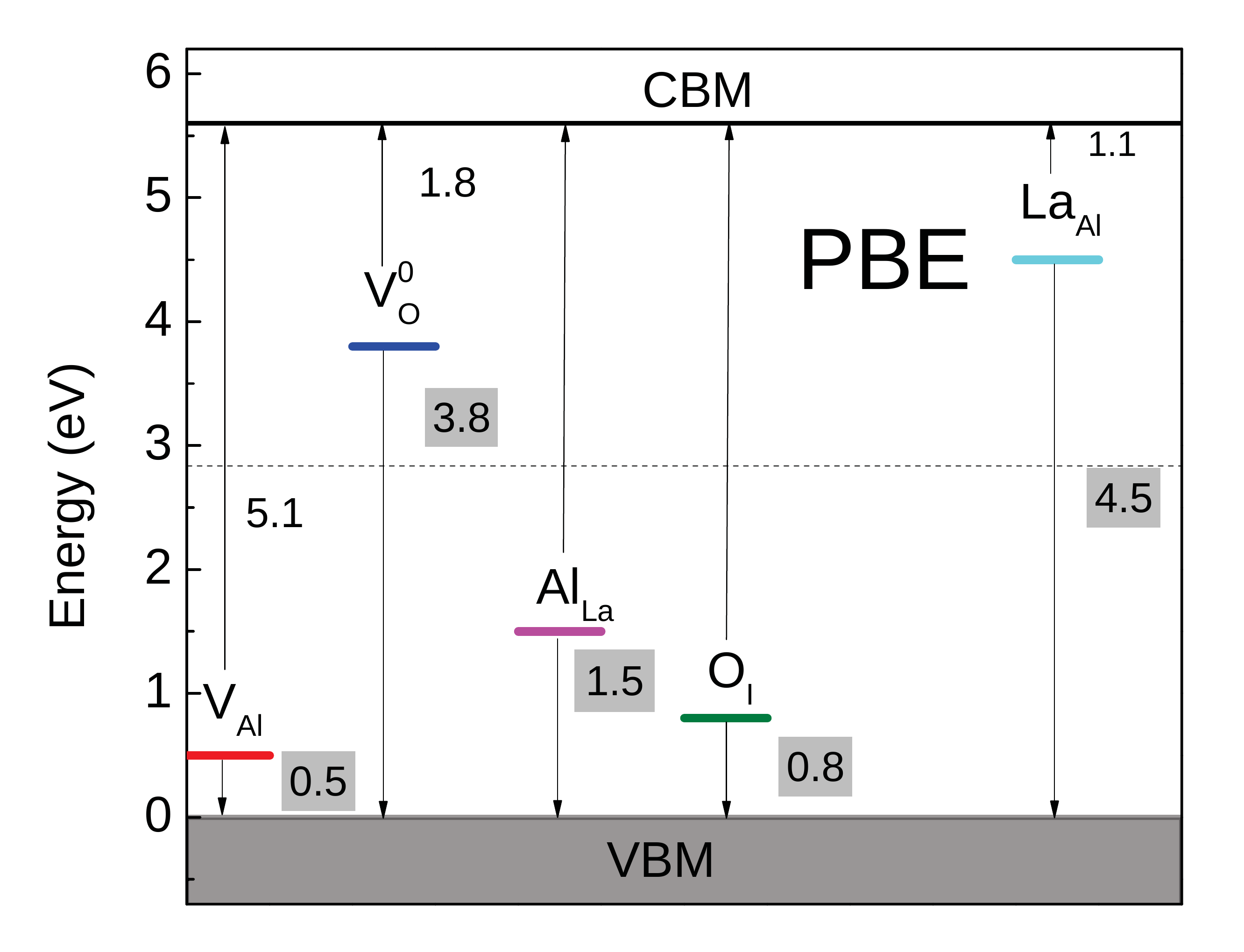}
\caption{\label{fig:defect} Schematic representation of  the average location of the defect bands in the band gap of LAO calculated with HSE/SZVP (left) and PBE from Ref.~\onlinecite{Luo:2009} (right) shifted using a scissor operator. Numbers in gray boxes refer to the location of the defect bands  with respect to the valence band maximum (VBM). The dashed line refer to the mid gap.}
\end{figure*}

The formation energies of  defects in LAO as function of its composition are plotted in Figure~\ref{fig:formation}.\cite{Note1} 
%~\footnote{ Point A: $\mathrm{\mu_O=\mu_O(bulk)}$, $\mathrm{ 2\mu_{Al}+3\mu_{O}=\mu_{Al_2O_3}(bulk)}$ where $\mathrm{\mu_O(bulk)}$ corresponds to the chemical potential per atom of O$_2$ gas.\\ Point B: $\mathrm{\mu_O=\mu_O(bulk)}$ , $\mathrm{ 2\mu_{La}+3\mu_{O}=\mu_{La_2O_3}(bulk)}$. \\ Point C: $\mathrm{\mu_{La}=\mu_{La}(bulk)}$, $\mathrm{ 2\mu_{La}+3\mu_{O}=\mu_{La_2O_3}(bulk)}$ \\ Point D:   $\mathrm{\mu_{La}=\mu_{La}(bulk)}$, $\mathrm{\mu_{Al}=\mu_{Al}(bulk)}$\\ Point E:   $\mathrm{\mu_{Al}=\mu_{Al}(bulk)}$,  $\mathrm{ 2\mu_{Al}+3\mu_{O}=\mu_{Al_2O_3}(bulk)}$}
Under oxidizing conditions (points A and B)  we identify the oxygen
interstitial (O$_\mathrm I$) as having  the lowest formation energy; this is
contrary to previous PBE results~\cite{Luo:2009} which predicted O$_I$ to be
less stable than \Vn{La} and other vacancy complexes.   \FE{It is worth noting that
we introduced the oxygen atom  at a random position in the supercell avoiding well-known interstitial sites followed
by a full relaxation of  the system. The resulting configuration
consists of a 110  split interstitial (dumbbell) with an O-O bond of
1.38~\AA. Since Luo {\it et al.};~\cite{Luo:2009} did not report their interstitial configuration, we could assume that our differences arise from different interstitial sites considered rather than computational. }

Focusing specifically at point A, \Vn{La} is the next most stable defect. Our formation
energy is about 3~eV  higher than previously published results obtained using the
PBE functional in rhombohedral and cubic LAO.~\cite{Luo:2009, Yamamoto:2012}  In
terms of competition between \Vn{La} and \Vn{Al}, we find (using HSE) the same
behavior seen using  PBE in Ref. ~\onlinecite{Chen:2011, Luo:2009}.  Next in
order of stability is 
\Vn{Al} and Al$_{\mathrm La}$ with equal formation energies at point A, followed
by \Vn{O}, a behavior not reported  previously.~\cite{Luo:2009, Yamamoto:2012}

Moving from point A to point B, the order of increasing stability of defect
types remain unchanged, except for Al$_{\mathrm La}$, which has become less
stable than \Vn{O}. We report a formation energy of 8~eV for \Vn{O}, which is
in excellent agreement with a recently computed HSE value of 8.3~eV in
rhombohedral LAO using a supercell of up to 135 atoms;\cite{Mitra:2012} note
that in this study other vacancy types and substitutions were not modeled

Under reducing conditions (point C, D, E), \Vn{O} dominates the spectrum, in
qualitative and quantitative agreement with previous uncorrected  PBE
calculations~\cite{Luo:2009},  having an average formation energy of 1.3 eV. \FE{ The formation energy of \Vn{O} calculated with HSE is lowered by 0.1~eV when the supercell size increases from 40 to 90 atoms.   Although not negligible, this remains smaller than the differences reported in the charged states~\cite{Luo:2009, Yamamoto:2012,Mitra:2012} which are due to  both the strong elastic and electrostatic self-defect interactions. Obviously, calculations using larger fully relaxed  supercells  are required to determine at what size defect self-interactions (elastic effects) become negligible. }

Our
results do not agree, however,  with the recent formation energies computed by
Yamamoto {\it et al.}\cite{Yamamoto:2012}  
who  applied a band gap correction (a 2.48~eV shift)  to
the PBE formation energies of \Vn{O}. Applying the  band gap correction in this case led
to the conclusion that Schottky-type vacancy complexes are more stable than
\Vn{O}. We believe this to be an artifact of the correction they applied.

It should be noted that interstitials
like La$_\mathrm I$ and Al$_\mathrm I$ are not addressed in the present study
because their neutral charge state was not identified to be stable according to
the PBE calculation of Luo {\it et al.}~\cite{Luo:2009} Also, our formation
energy spectrum computed with HSE reveal that they exhibit very high formation
energies. 

%\FE{Overall, there are some quantitative differences between our HSE results and previous uncorrected  PBE results~\cite{Luo:2009}. These might originate from the enhanced accuracy we gain with HSE in the calculated valence band width (VBW) and band gap of bulk \STO~(see table~\ref{tab:VBW}) identified in Ref. \onlinecite{Ramprasad:2012} to reflect an enhanced precision in the defect formation energies.}

%%%%%%%%%%%%%%%%%%%%%%%%%%%%%%%%%

The various defects we will first discuss induce changes to the  electronic
properties of cubic  \LAO, introducing defect levels within the band gap and/or
lifting the degeneracy of  the CBM and VBM as shown in figure~\ref{fig:bands}.

The oxygen split interstitial configuration (O${_I}$), which is the most stable
under oxidizing conditions, induces a strong distortion to the lattice, which in turn
significantly impacts the electronic structure. The CBM splits at $\Gamma$
point by 440~meV, while the VBM also splits into three distinct bands. The fully
occupied defect band composed of O 2$p$ states is located on average at 0.13~eV
above the VBM, has valence band character, and induces a gap of 4.66~eV. 
\Vn{La}, the second most stable defect under oxidizing conditions, creates three
empty non-degenerate valence bands, dominated by O 2$p$ orbitals originating
from the O dangling bonds.  Both HSE and PBE agree about the nature and the
location of these bands. However, our O${_I}$ level is shallower than the
previously reported PBE results,~\cite{Luo:2009} which is probably due to
differences in the interstitial configuration.

Next to be evaluated are defects having in-gap states, namely Al$_{\mathrm La}$,
\Vn{O}, La$_{\mathrm Al}$ and  \Vn{Al}, which show a localized electronic
density around the defect region. The Al$_{\mathrm La}$ antisite defect might play a role under  oxidizing conditions
due to its relatively low formation energy. With HSE, we find that it induces an
empty defect band in the gap at 2.93~eV above the VBM and 2.0~eV below the CBM.
This band might become populated upon doping or under excitation, and 
is dominated by O 2$p$ and Al $s$ orbitals ({\it q.v.} the  PDOS). The  bulk
degeneracy of the VBM and CBM are  not affected,  and remain 3 and 2 fold
degenerate, respectively. This is an indication that this defect  does not
introduce noticeable distortion or octahedral rotation into the lattice, which
is further confirmed by a  structural analysis. However, using  PBE we find that
the  Al$_{\mathrm La}$  defect band is located at 1.22 eV above the VBM and 2~eV
below the CBM, which is well below the mid gap (1.6~eV). Following a typical 
band gap correction procedure,  this PBE defect band does
not need to be shifted using the scissor operator,~\cite{Luo:2009} which would result in keeping its  VB character, which contradicts the HSE results above.

The next defect of interest is \Vn{O}, which is arguably the most important
defect under reducing conditions, and suspected to be systematically
introduced during the growth of metal oxide superlattices.~\cite{Yao:2013} After
introducing \Vn{O}, the supercell shrinks along the $y$ axis, leading to a
tetragonal distortion of the lattice with   a ratio $a/b$=1.0057 ($a$ and $b$
being the new lattice parameters) and a slight rotation of \alo octahedra. This
impacts strongly the electronic structure by splitting the doubly degenerate CBM by
258~meV, while leaving the VBM triply degenerate.  A new defect band also appears
at 2.77~eV above the VBM from the  combination of O 2$p$, Al $d$, La $d$ and $p$
orbitals. Here again, major differences with previous PBE data emerges: the
uncorrected PBE level computed recently by  Chen {\it et al.}~\cite{Chen:2011}
was located 2.23 eV above the VBM.  Luo {\it et al.} applied the
scissor operator to this defect level, predicting it to lie at about 3.8~eV above
the VBM. 

Last to be examined is La$_{\mathrm Al}$, which in the neutral state would
rarely form under either oxidizing or reducing conditions. It introduces a 
fully-occupied triply degenerate defect band located 2.06~eV above the VBM and 2.60~eV
below the CBM. However, the PBE defect level is at that method's mid gap,
lying 1.6~eV from the VBM and CBM.  If a scissor operator was to be used, one could argue that this level should be shifted, placing it as close as 1~eV to the CBM (see
figure~\ref{fig:defect}). 

To conclude, there are fundamental differences between our HSE defect level
spectrum and the one published earlier using corrected  PBE~\cite{Luo:2009}
data regarding the nature of the defect bands (see figure~\ref{fig:defect}).
We believe these differences originate from the criterion used to judge whether
the ``scissor operator'' should be applied. For example,  HSE finds that
Al$_{\mathrm La}$ and V$_{\mathrm Al}$ have defect bands near mid gap, thus
removing the PBE's prediction of valence band character, which were reported
previously. The same issue leads to significant differences in the conclusions
regarding \Vn{O}. Overall, our defect levels calculated with HSE lie 2~eV below
the CBM (see figure~\ref{fig:defect}),  which is in better agreement with
recent experiment~\cite{Chen:2011}.  This HSE defect level spectrum we propose
here  is correction free, and may be used to interpret experimental
photoluminescence data which place defect levels at 3.1, 2.1 and 1.7
eV.~\cite{Liu:2011, Chen:2011} 

% According to our diagram the 600 nm peak~\cite{Chen:2011} could originate from either the O$_I$,  V$_{La}$ and other charged defects not considered in this work. The Al$_{La}$ antisite  defect level is the highest but still lies at 2~eV below the CBM and might be  at the origin of the fast decay process observed experimentally. 2.9 eV compared to 3.1~eV. Vo is and LaAl the 2.1 finally VAl the 1.7 eV.

% \\\\\\\\\\\\\\\\\\\\\\\\\\\\\\\\\\\\\\\\\\\////////////////////////////////////////////////////////
%: ||||||||||||||||||||||||||||||||||||   ACKNOWLEDGMENTS    |||||||||||||||||||||||||||||||||||||||||||||||||||||||| 
\begin{acknowledgments}

This work is supported by the  Qatar National Research Fund  (QNRF) through the National Priorities  Research Program (NPRP  08 - 431 - 1 - 076).  GES acknowledges support from The Welch Foundation (C-0036).  We are grateful to the research computing facilities at Texas A\&M University at Qatar for generous allocations of computer resources.\\

\end{acknowledgments}
%\bibliography{LAO-2013bb}

\end{document}